\documentclass[namedreferences]{kluwer}

\usepackage{graphicx}

\def\fig{.}
\def\dd{{\rm d}}

\begin{document}

\begin{article}

\begin{opening}

\title{  
The subsurface radial gradient of solar angular velocity
        from MDI f-mode observations
        }


\author{T. \surname{Corbard}}
\author{M.~J. \surname{Thompson}}
\institute{Space and Atmospheric Physics Group, The Blackett Laboratory, 
Imperial College, London SW7 2BW, UK}
             
\runningauthor{Corbard \& Thompson}
\runningtitle{The subsurface radial gradients}
\begin{abstract}
 We report quantitative 
analysis of the radial gradient of solar angular velocity at depths down to
about 15 Mm below the solar surface for latitudes up to $75^\circ$ 
using the Michelson Doppler Imager (MDI) observations of surface gravity waves
(f modes) from the Solar and Heliospheric Observatory (SoHO). 
A negative outward gradient of around $-400$nHz$/R_\odot$, equivalent to 
logarithmic gradient of the rotation frequency with respect to radius
which is very close to $-1$, is found to be remarkably 
constant between the equator 
and $30^\circ$ latitude. Above $30^\circ$ it decreases in absolute magnitude
to a very small value at around $50^\circ$. At higher latitudes the gradient 
may reverse its sign: if so this reversal takes place in a thin layer 
extending only 5 Mm beneath the visible surface, as evidenced by the
most superficial modes (with degrees $l>250$).
The signature of the torsional oscillations is seen in this layer, but
no other significant temporal variations of the gradient and 
value of the rotation rate there are found.
\end{abstract}  

\end{opening}

\section{Introduction}

The velocity field of the rotational flow in the Sun's near-surface layers 
may play a significant role in small-scale dynamo action in that region and
in the dynamics of supergranular convection. Surface observations over 
decades and even centuries have shown that the latitudinal variation of 
the surface rotation is rather smooth, being rather well described by 
a three-term (i.e. second-order) polynomial in 
$\mu^2$ where $\mu=\cos\theta$ and $\theta$ is the colatitude.
Recent analyses of high-resolution data, 
in particular those utilizing solar f-mode observations by the Michelson Doppler
Imager (MDI) 
on board the Solar and Heliospheric Observatory (SoHO), have highlighted important
departures from such a description of the rotation of the near-surface layers.
The polar subsurface layers (i.e. $\theta<20^\circ$ and  depths down to $28$Mm 
below the surface) 
have been shown to be approximately $10$nHz slower than expected from
a simple three-term extrapolation from lower latitudes
\cite{birch98,schou:mdi98,schou99}; and   
\inlinecite{koso_schou97} have shown that, 
at a depth of 2 to 9 Mm beneath the surface,
there exist zonal bands of 
alternate faster and slower rotation rate of $\sim \pm 5$m/s 
superimposed on the general trend described by the second order polynomial.
This latter feature, inferred from the first observations of MDI in 1996, 
was found to be similar to the surface `torsional oscillations' 
\cite{howard80} also observed in 1995 in Doppler measurements using the 
first GONG observations \cite{hathaway96}.) More recently still, 
analysis of both p and f modes from the GONG network 
and MDI instrument have further 
led to the conclusion that these banded structures
extend at least down to $60$Mm below the surface \cite{howe2000_band}. 

The observed f modes, being confined to the outer layers of the Sun, provide 
a relatively clean and straightforward measure of conditions there.
But those results above that were obtained just from the f modes
assumed at least implicitly that the angular velocity is not 
varying significantly with depth within the layer sensed by 
those modes. 
It is however well known that another important property of the subsurface 
layers is that 
they present a radial gradient of angular velocity. This was first
suggested by the fact that different indicators 
such as Doppler shifts of photospheric Fraunhofer lines, various magnetic field
features of  different ages and sizes (sunspots, faculae, 
network elements, $H_\alpha$ filaments) or
the supergranular network, present different rotation rates 
(see the review of \opencite{howard84}; \opencite{schroeter85}; 
\opencite{snodgrass92}). 
This has been interpreted by assuming that the different 
magnetic features are anchored at different depths
(e.g. \opencite{foukal72}; \opencite{collin95}), their different 
rotation rate being therefore interpreted as an indication 
of the existence of radial gradients of 
angular velocity in the subsurface layers.
More specifically, noticing
that the supergranular network  rotation 
rate ($\sim 473$nHz) was found to be $\sim 4\%$ faster than the upper
photospheric plasma rate 
obtained from spectroscopic methods and also
$\sim 2\%$ faster than various  magnetic indicators  thought to be rooted
under the supergranulation layer, \inlinecite{snodgrass_ulrich90}  
inferred that a maximum of angular velocity should exist somewhere 
between $0.95R_\odot$ and the surface.

{}From the theoretical point of view, it has been suggested that the 
angular momentum per unit mass $\Omega r^2 \sin^2\theta$ 
could be conserved in the supergranular flow
 \cite{foukal_jokipii75,foukal77,gilman_foukal79}.
{}From  $\partial\Omega/\Omega =-2\partial r/r$, at fixed latitude, this simple argument 
leads effectively to a negative gradient below the surface, and 
the $4\%$ difference 
in rotation rates would be explained if the supergranulation network velocity 
observed at the surface were reflecting
the rotation rate at a depth of $2\% R_\odot\simeq 15$Mm,
which turns out to correspond to the depth expected for the 
supergranular convection \cite{foukal77,duvall80} (but see also \inlinecite{beck00} 
for more recent estimate).
In order to reproduce the observed patterns of solar activity such as the 
equatorward migration of sunspots, early dynamo models based on a positive 
surface $\alpha$-effect
indicated also that the angular velocity  must decrease outwards i.e. 
$\partial\Omega/\partial r<0$ 
(e.g. \opencite{leighton69}; \opencite{roberts_stix72}).
One of the first goals of helioseismology was therefore to test the 
assumptions about the negative gradient
of angular velocity below the surface suspected from different surface 
observations. 
This was indeed first attempted 
by \inlinecite{Deubner79}: although they did not resolve individual modes, they 
were able, from ridge-fitting separately the eastward- and westward-propagating
near-equatorial waves in the ($k$, $w$) diagram, to detect such a 
negative gradient close to the surface. 
Subsequent helioseismic work using resolved mode frequencies
has shifted much theoretical focus to the base of the 
convection zone by showing that the radial gradient of angular velocity
in the bulk of the convection zone is weak and that 
a strong radial shear, the so-called tachocline, occurs 
at its base. The gradient $\partial\Omega/\partial r$ is positive in the 
tachocline at sunspot (i.e. low) latitudes \cite{brown89}. 
This has led various dynamo theories to locate 
the dynamo action below the convection zone,
 with a negative $\alpha$-effect operating there 
(e.g. \opencite{gilman89}; \opencite{parker93}) though some recent work has 
revisited the idea of a
positive surface $\alpha$-effect but invoking
the action of a meridional circulation, 
equatorward below the convection zone and poleward at the surface,
to produce the observed equatorward migration of sunspots by advective 
transport of flux \cite{dikpati_char99,Kuker01}.
The lack until recently of precise determinations of high-degree mode 
parameters made it
difficult to obtain very localized inferences about 
the subsurface layers. But, because all the observed 
modes have amplitude close to the surface, 
inverters again got hints about the existence
of a radial shear close to the surface (especially using methods such as
regularized least-squares which readily extrapolate into regions where
the data provide no localized information) though without being able to 
quantify precisely its extent and
amplitude (e.g. \opencite{thompson:gong96}; \opencite{corbard97}). 
   
We show in this work that f-mode observations allow us to make 
quantitative inferences about the surface radial shear. These should be
taken into account when modeling near-surface dynamo action 
or the dynamics of the supergranulation layer.

\section{Observations}
The data used here are 23 independent
times series of 72 days obtained from the so-called MDI medium-$l$
program. These cover the period from 1996 May 1 to 2001 April 4 with 
interruptions 
during the summer 1998 (June 23 to October 23) and between 1998 December 4 and 
1999 February 4 due to SoHO spacecraft problems. More details on 
the production of these time series from the observations 
can be found in \inlinecite{schou99}.

A given f-mode multiplet in the spectra comprises
$2l+1$ frequencies $\nu_{lm}$, where $l$ and $m$ are the degree and 
azimuthal order of the spherical harmonic $Y_l^m(\theta,\phi)$ describing
the angular dependence of the modes.
The so-called $a$ coefficients for the multiplet are defined
by the polynomial expansion:
\begin{equation}\label{eq:fit}
\nu_{l m}=\nu_{l 0}+\sum_{j=1}^{2l} a_j^l {\cal P}_j^{(l)}(m) \ \ m=\pm 1,\pm2...\pm l
\end{equation}
 where ${\cal P}$  are orthogonal polynomials normalized such that
 ${\cal P}_j^{(l)}(l)=l$ \cite{schou94}. 
All f modes considered here have degrees between $l=117$ and $l=300$ but 
the total number of multiplets observed is between 112 and 143, depending  
on the 72-day interval considered.    
For each observed mode, the central frequencies $\nu_{l 0}$ and the 
first $36$ $a$ coefficients have
been estimated using the method described in \inlinecite{schou_thesis}.
Odd-indexed $a$ coefficients, which describe the dependence of the frequencies
that is an odd function of $m$, arise from
the North-South symmetric part of the solar rotation.
Even-indexed coefficients 
arise from latitudinal structural variation, centrifugal
distortion and magnetic fields.

\section{Data analysis }

Following  \inlinecite{ritzwoller_lavely}, 
we identify the North-South symmetric part of the angular velocity  
$\Omega(r,\mu)$ with the odd-degree, zonal part of the 
toroidal component of a general stationary and laminar velocity field 
and write 
\begin{equation}
\label{eq:omega}
\Omega(r,\mu)=\sum_{j=0}^\infty \Omega_{2j+1}(r) 
\bar{T}^1_{2j}(\mu)\;,
\end{equation}
where $r$ is fractional radius  and 
$\bar{T}_{2j}^1\!\!\equiv\!\!{T_{2j}^1(\mu)}/{T_{2j}^1(0)}$ 
 are Gegenbauer polynomials (see Appendix)  normalized 
 such that the equatorial rate is given by the straight sum of the 
$\Omega_{2j+1}(r)$.

Assuming slow rotation, we can use a linear perturbation 
theory to predict the effect of rotation on the oscillation modes 
(e.g. \opencite{hansen77}). Moreover, with the 
polynomials $\cal P$ and expansion Equation~(\ref{eq:omega}) as chosen,
there is a one-to-one 
relation between odd $a$ coefficients and the components 
 $\Omega_{2j+1}(r)$  \cite{ritzwoller_lavely}, thereby reducing the full 
2D problem to a set 
of 1D integral equations often refered as the 1.5D problem. In the particular 
case of the f modes, we obtain
\begin{equation}\label{eq:15D}
2\pi a^{l}_{2j+1}= u_{2j+1}^l \int_0^1 \!\!K_h^l(r)
\Omega_{2j+1}(r)\dd r \;,
\end{equation}
where the expression for  the kernels $K_h^l(r)$ and  $u_{2j+1}^l$ are derived 
in Appendix.

\begin{figure}[ht]
\resizebox{\hsize}{6cm}{\includegraphics[angle=90]{{\fig}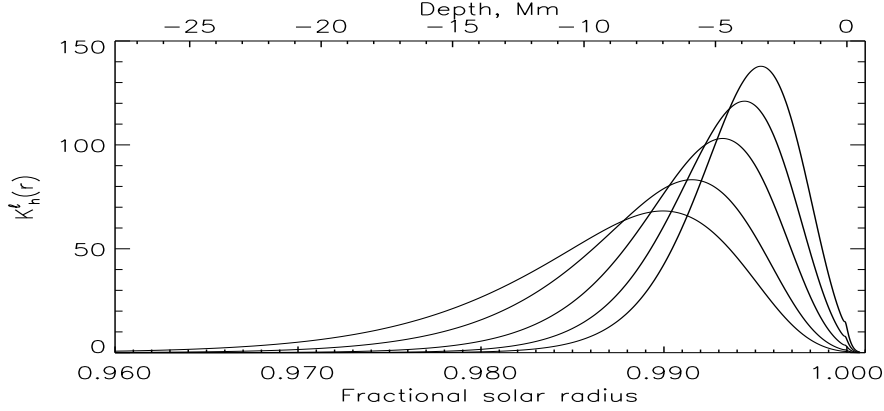}}
\caption{f-modes rotational kernels $K_h^{l}(r)$ for $l$=117, 150, 200, 250, 300 from left to right.}
\label{fig:kernels}
\end{figure}
The 36 $a$ coefficients extracted from 
observation do not provide information about the terms above $j=17$ 
in the summation in  Equation~(\ref{eq:omega}) and that corresponds to a 
limitation in the latitudinal resolution we can reach. 
Defining
\begin{equation}
\label{eq:15Dcoefs}
b_{2j+1}^l\equiv\frac{2\pi a_{2j+1}^l}{u_{2j+1}^l }\;,
\end{equation}
from Equations~(\ref{eq:omega}), (\ref{eq:15D})  and (\ref{eq:15Dcoefs}) we obtain
\begin{equation} \label{eq:fundamental}
\sum_{j=0}^{17} b_{2j+1}^l \bar{T}_{2j}^1(\mu_0)
\approx\int_0^1{K_h^{l}(r)}{\bar\Omega(r,\mu_0)}\dd r\;,
\end{equation}
where $\bar\Omega(r,\mu)$ refers to the part of the rotation profile 
corresponding to the sum Equation~(\ref{eq:omega}) truncated at $j=17$.
\begin{figure}[ht]
\resizebox{\hsize}{4cm}{\includegraphics[angle=90]{{\fig}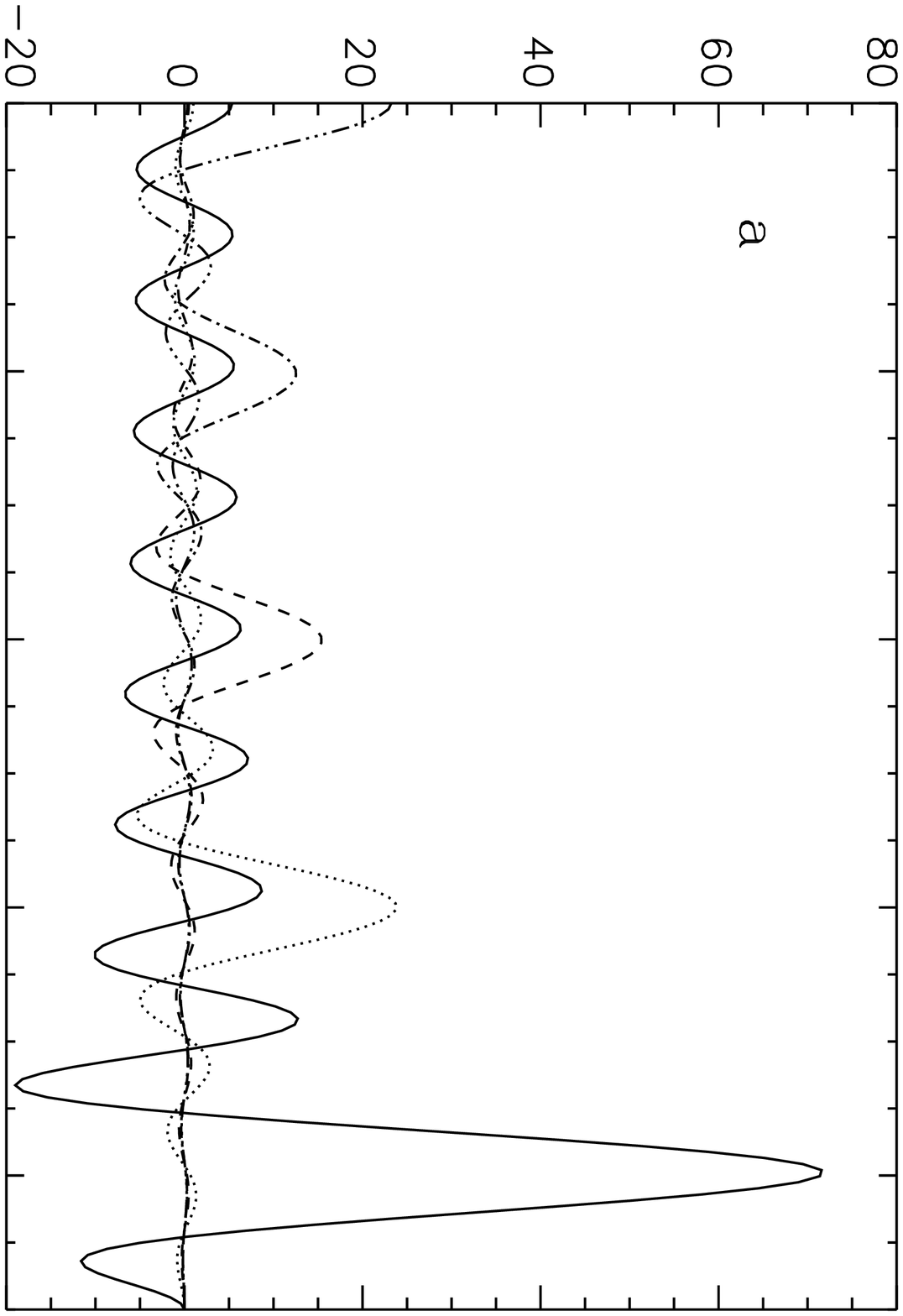}}
\vskip -.5cm
\resizebox{\hsize}{4cm}{\includegraphics[angle=90]{{\fig}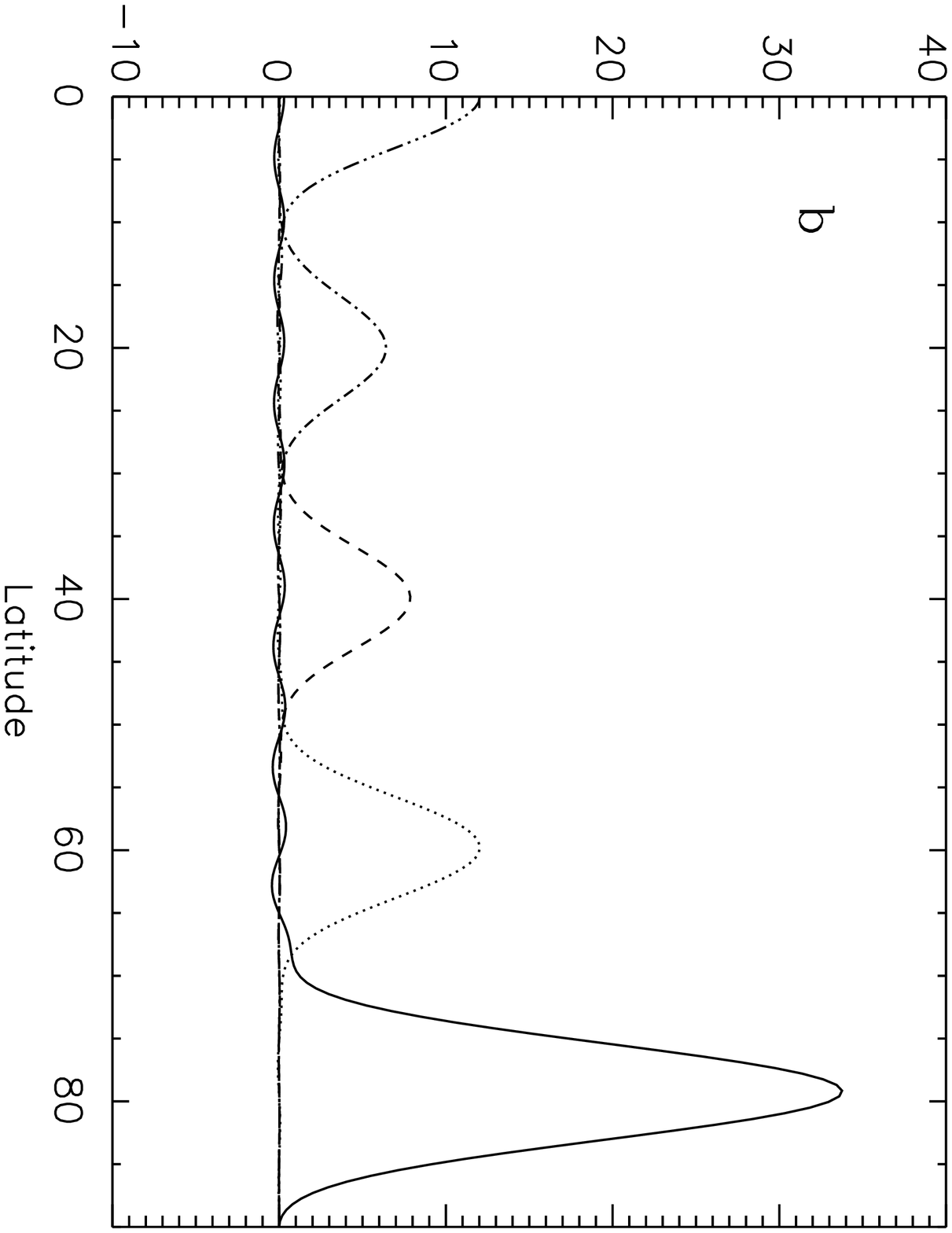}}
\vskip .5cm
\caption{Latitudinal averaging kernels at 0, 20, 40, 60, $80^\circ$ of latitude (double-dot-dash, dot-dash, dash, dot and full lines respectively) corresponding to the combination  {\bf a} Equation~(\ref{eq:15D_bis}), {\bf b} Equation~(\ref{eq:117}).}
\label{fig:kern_lat}
\end{figure}
We can also show \cite{pijpers97} that 
the above linear combination of $b$ coefficients is such that
\begin{equation}\label{eq:av_kern}
\sum_{j=0}^{17} b_{2j+1}^l \bar{T}_{2j}^1(\mu_0) = \int_0^1 \int_0^1 K_h^l(r) \kappa(\mu_0,\mu)
\Omega(r,\mu)\dd r \dd\mu\;,
\end{equation}
where $\kappa(\mu_0,\mu)$ are the so-called latitudinal averaging kernels 
which show what latitudinal average of the true rotation rate is made  
at each latitude. 
Figure.~\ref{fig:kern_lat}a shows that these kernels have their main
peak centered at $\mu_0$ but present an oscillatory behaviour which may 
lead to systematic errors if some small-scale features (corresponding to terms
with $j>17$) exist in the true rotation rate. In order to avoid this, 
one may try to find instead the combination of $b$ coefficients that leads 
to kernels that are optimally localized around a given latitude. This can be 
achieved following for instance the method of \inlinecite{backus68},
but we notice here that a similar result can be obtained simply by 
introducing, in the sum of  Equation~(\ref{eq:av_kern}),
a correcting factor $e^{-j(j+3/2)/l_0}$  where $l_0\equiv117$ corresponds to 
the lowest degree of the observed f modes (see also Equation~(\ref{eq:v_alt})). 
Doing this, the latitudinal 
averaging kernels are found better peaked (Figure~\ref{fig:kern_lat}b)
and the formal errors associated 
with the linear combination of the $b$ coefficients is lowered.
Following the definition of \inlinecite{corbard_tacho_rev}, the latitudinal 
resolution obtained is about $10^\circ$ at all latitudes.

The kernels ${K_h^l(r)}$ associated with each f mode have a simple
shape with only one maximum located at slightly different radial
positions depending on the degree $l$ (Figure~\ref{fig:kernels}).
If we define $r_0^l\equiv\int_0^1{K_h^l(r)r\dd r}$, the radial location of
the center of gravity of these kernels, 
and assume a linear behaviour of the rotation rate at each latitude in the 
radial domain where the f modes considered have appreciable amplitude, i.e.,
\begin{equation}\label{eq:lin_fit}
\Omega(r,\mu_0)\!=\!\alpha(\mu_0)-\beta(\mu_0)(r-1)
\end{equation} 
in $r > 0.97$, say,
we simply obtain
\begin{equation}
\label{eq:15D_bis}
 \sum_{j=0}^{17} b_{2j+1}^l \bar{T}_{2j}^1(\mu_0) \approx \bar\Omega(r_0^l,\mu_0)\;,
\end{equation}
where the meaning of $\bar\Omega$ is the same as in Equation~(\ref{eq:fundamental}).
Alternatively, a slightly modified choice of weights yields
\begin{equation}\label{eq:117}
 \sum_{j=0}^{17} b_{2j+1}^l \bar{T}  _{2j}^1(\mu_0) e^{-\frac{j(j+3/2)}{117}}
\approx <\!\!\Omega(r_0^l,\mu)\!\!>_{\mu_0}
\approx \Omega(r_0^l,\bar\mu_0)\;,
\end{equation}
where the brackets denote the weighted average around $\mu_0$, the 
weighting function being the kernels of Figure~\ref{fig:kern_lat}b. The second 
approximate equality in Equation~(\ref{eq:117})
would be exact if the rotation profile were
a linear function of $\mu^2$ in the domain covered by 
the averaging  kernels (i.e. $\pm 10^\circ$), with 
$\bar\mu_0^2\equiv \int_0^1 \kappa(\mu_0,\mu) \mu^2 \dd\mu$; the
approximation is less good, however, at high solar latitudes.

The parameters $\alpha$ and $\beta$ can then be estimated at each latitude 
from a linear least-square fit, yielding not only an estimate of the 
value of the rotation rate at e.g. the surface, but also an estimate of
the average gradient $\partial\Omega/\partial r$ in the region sampled
by the f modes. Finally we note that the dependence of $\Omega$ as a function
of radius in the near-surface layers may sometimes conveniently be 
described by a power of $r$: we note that this description
is immediately derivable from our
$\alpha$ and $\beta$, since for small values of $1-r$ the right-hand 
side
of Equation~(\ref{eq:lin_fit}) is well approximated by 
$\alpha(\mu_0) r^{-\alpha(\mu_0)/\beta(\mu_0)}$.

\section{Results}

\begin{figure*}
\resizebox{\hsize}{8cm}{\includegraphics[angle=90]
{{\fig}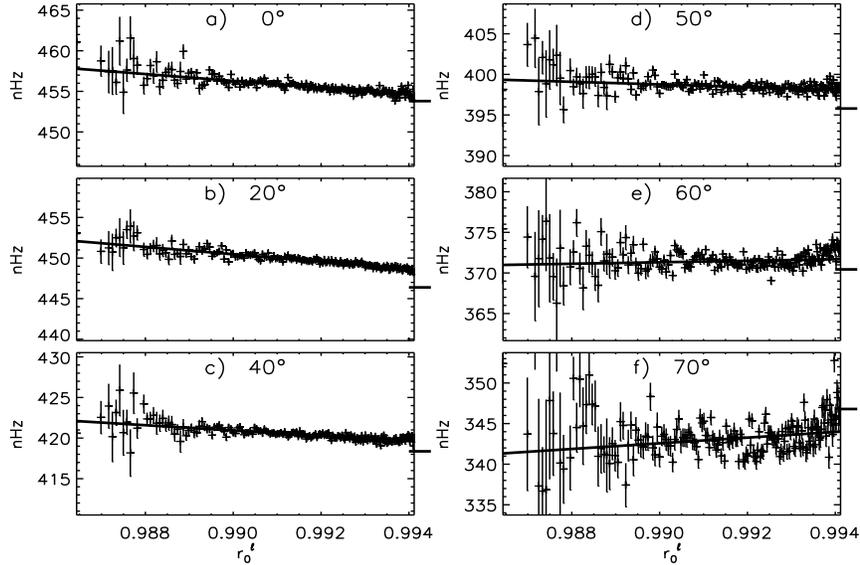}}
\caption{
Time average of 
$\bar\Omega(r_0^l,\mu_0)/2\pi$
(Equation~(\ref{eq:15D_bis})) for values of  $\mu_0$ 
corresponding to the latitudes indicated in each panel.  
The result of the linear fits (Equation~(\ref{eq:lin_fit})) are shown by the 
straight lines. The error bars are the standard 
deviation associated with the weighted temporal mean. The mark on the right 
of each panel indicate the surface plasma rate obtained by 
Snodgrass et al. (1984).
Note that the surface spectroscopic value indicated on panel f is 
essentially an extrapolation from observations at lower latitudes.
}
\label{fig:fit_lin}
\end{figure*}

\begin{figure}[ht]
\resizebox{\hsize}{10cm}{\includegraphics[angle=90]
{{\fig}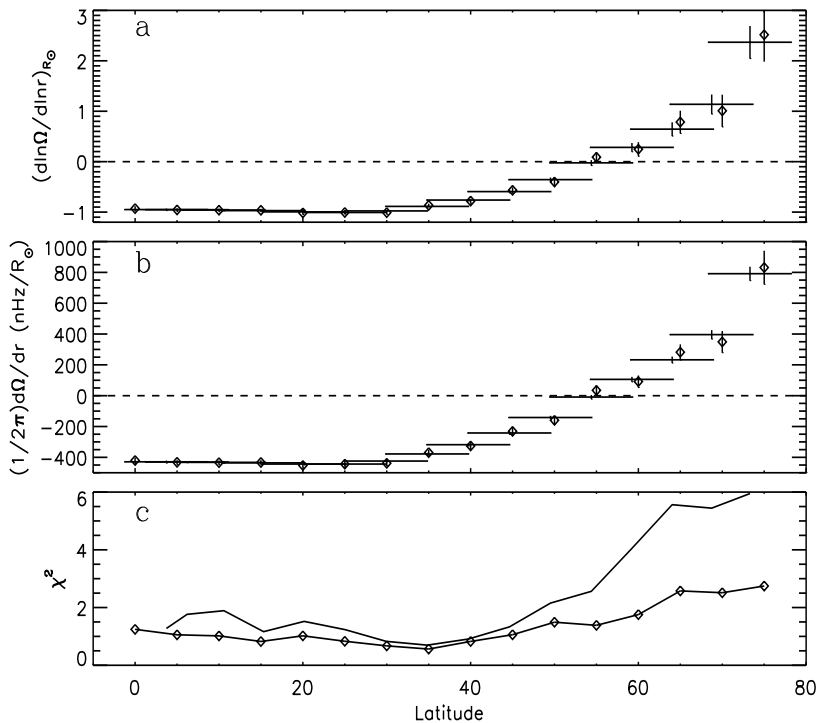}}
\caption{{\bf a} Logarithmic derivative of angular velocity as a function 
of latitude. This corresponds to the ratio $\beta/\alpha$ of Equation~(\ref{eq:lin_fit}). {\bf b} Radial gradient of angular velocity $\beta$ as a function of latitude.
 {\bf c} Normalized $\chi^2$ value of the linear fit. The diamond symbols
are for the results obtained using Equation~(\ref{eq:15D_bis}) while 
the other points are obtained using Equation~(\ref{eq:117}). 
The horizontal error bars indicate the angular resolution as deduced 
from Figure~\ref{fig:kern_lat}b. The vertical error bars are formal errors 
deduced from the linear fit. The dashed horizontal line correspond to no 
radial gradient of angular velocity.} 
\label{fig:ab-b-chi2}
\end{figure}

By combining the frequency splittings within each f-mode multiplet 
in the manner given by Equation~(\ref{eq:15D_bis}), for different choices
of target latitude, we obtain measures of the near-surface rotation 
which are reasonably well localized in latitude and which correspond to
different weightings in the depth direction. The latitudinal sensitivity
is illustrated in Figure~\ref{fig:kern_lat} and the depth sensitivity in 
Figure~\ref{fig:kernels}.
Figure~\ref{fig:fit_lin} shows the results of combining the data 
using Equation~(\ref{eq:15D_bis}),
averaged in time over all the datasets under study. In 
depth, the points are plotted at the center of gravity ($r=r_0^l$)
of the corresponding kernels (cf.~Figure~\ref{fig:kernels}). 
It is evident from these
results that, at low latitudes, the weighted rotation increases with depth. 
If at each latitude separately 
we fit these results to a rotation profile that is 
linear in depth, we obtain the linear fits overplotted in 
Figure~\ref{fig:fit_lin}. These
provide an average rotational gradient $\beta(\mu_0)$
in the outer 15 Mm or so
of the solar interior, and an extrapolated surface rotation rate
$\alpha(\mu_0)$. The gradient, as a function of latitude, is 
presented in Figure~\ref{fig:ab-b-chi2}, both in terms of its dimensional 
value and in terms of the logarithmic derivative
$\partial\ln \Omega / \partial\ln r$. 
It may be seen that 
for latitudes below $50^\circ$ the gradient of rotation with depth is
negative; at about $50^\circ$ it is close to zero; and for higher latitudes
the average rotational gradient becomes positive. We note that the
radial gradient is remarkably constant at latitudes up to $30^\circ$, and 
the value of the logarithmic derivative at these latitudes is close to 
$-1$. We return to this in the Discussion.

\begin{figure}[ht]
\resizebox{\hsize}{!}{\includegraphics[angle=90]
{{\fig}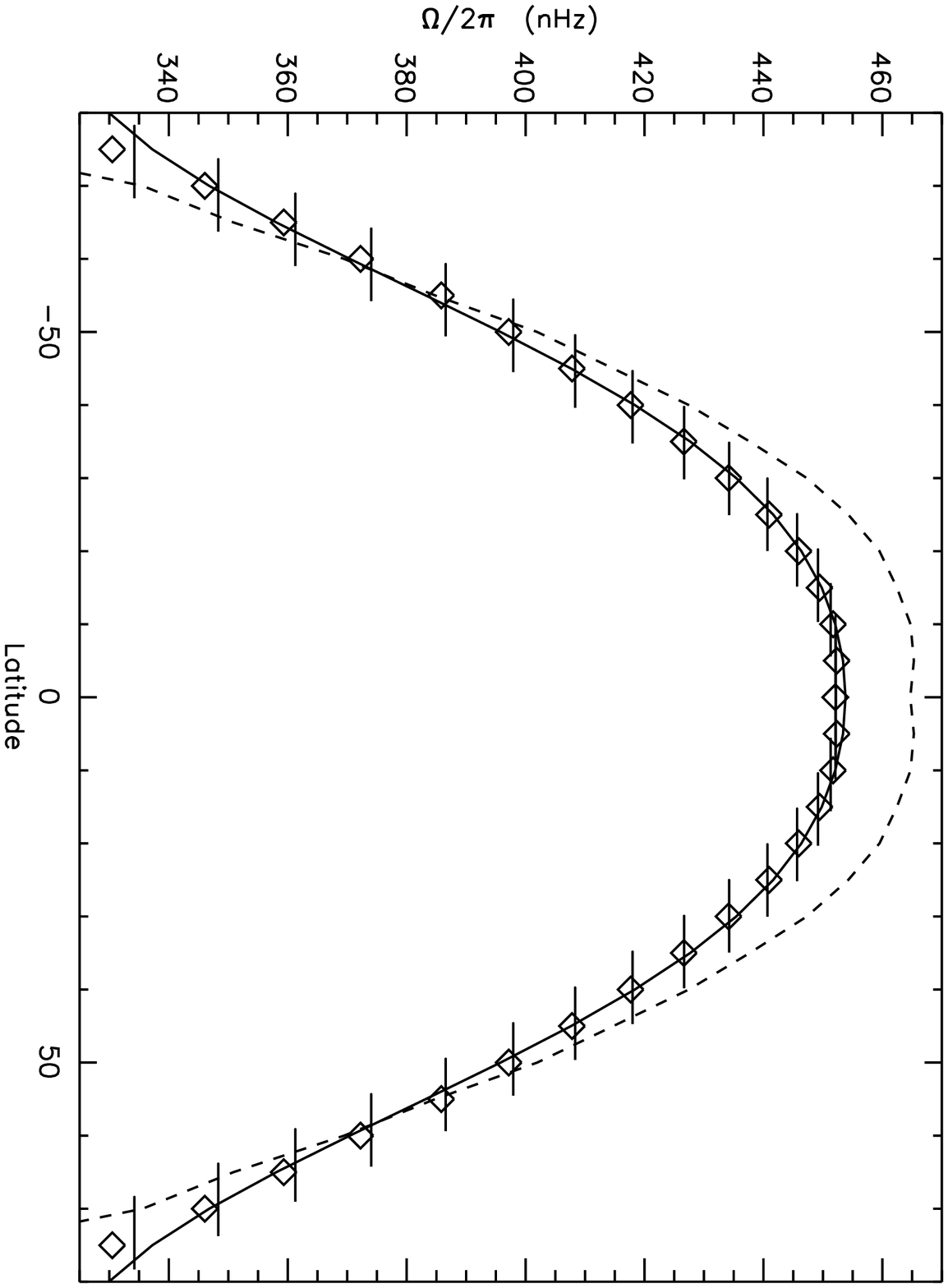}}
\caption{
The full line gives the photosheric plasma rotation rate inferred by 
Snodgrass et al. (1984); the diamond symbols and horizontal bars correspond
to $\alpha$, the intercept of the linear fit respectively in the case of 
Equation~(\ref{eq:15D_bis}) and Equation~(\ref{eq:117}) 
and the dashed line correspond to 
an extrapolation of the rotation rate at $0.97R_\odot$ using Equation~(\ref{eq:lin_fit}) 
in the case of Equation~(\ref{eq:15D_bis}).}
\label{fig:surface}
\end{figure}
Another way to visualize the changing gradient with latitude is that
in Figure~\ref{fig:surface}, where we show the rotation rate 
extrapolated both to the
surface ($r=1$) and to $r=0.97$. The deeper rotation is faster than the
surface rotation at low- and mid-latitudes, but slower at high latitudes.
At low- and mid-latitudes the extrapolated surface rate agrees well with
the spectroscopic surface measurements, given the approximately $1.5\%$
spread in recent such determinations 
(see the review by \inlinecite{beck_review00}). 
For comparison, we have 
made a fit to our inferred surface rate below 
$60^\circ$ latitude and present our fitting coefficients with the
spectroscopic coefficients of \inlinecite{snodgrass_HW84} in Table~\ref{tab:table1}. 
Similarly to what has been found previously, our inferred rotation rate
above $70^\circ$ is markedly slower than what would be expected from a 
3-term fit at low- and mid-latitudes: we return to this issue of the 
so-called `slow pole' later.

\begin{table*}[ht]
\caption{Comparison between the surface plasma rate  
 and our results from f modes analysis.}
\label{tab:table1}
\begin{tabular}{lccc}
\hline
 &$\Omega_1/2\pi$ & $\Omega_3/2\pi$  & $\Omega_5/2\pi$ \\
Method  &(nHz)&(nHz)&(nHz)\\
\lcline{1-1}\rlcline{2-2}\rlcline{3-3}\rlcline{4-4}
\inlinecite{snodgrass_HW84}\tabnote{Spectroscopic measurements made at the Mount Wilson 150-foot Tower
between 1967 and 1982.} & 436.4 & 21.0 & -3.6 \\
f modes ($l$-averaged)\tabnote{Average of the first 3 b 
coefficients (cf. Equations~(4), (8)).}
 & & & \\
\ \ ($117\le l \le 300$) \hfill $<r_0^l>=0.991$\tabnote{Center of gravity of the 
corresponding $l$ averaged radial kernels.} & 438.8 & 21.0 & -3.9 \\
\ \ ($160\le l \le 250$) \hfill $<r_0^l>=0.991$ & 438.9 &  21.2 & -4.0 \\
f modes (surface extrapolation)\tabnote{Obtained 
by fitting the intercept $\alpha(\mu)$  to the 
expansion Equation~(2)
for latitudes below $60^\circ$.} & & & \\
\ \ ($117\le l \le 300$) & 435.8 & 20.2 & -3.2 \\
\ \ ($160\le l \le 250$) & 435.7 &  20.5 & -3.6 \\
\hline
\end{tabular}
\end{table*}

Figure~\ref{fig:ab-b-chi2}c shows the chi-squared for the least-squares
fits at each latitude. 
The large chi-squared values at higher latitudes are striking. 
The difference between the chi-squared values when
using equations (\ref{eq:15D_bis}) and (\ref{eq:117}) is also very noticeable:
this arises largely
because the error bars on the fitted points are
reduced by the exponential factor in Equation~(\ref{eq:117}), which results in 
an increased chi-squared. Thus the interpretation of the
absolute value of the chi-squared may be a little uncertain, but the 
trend with latitude for the two cases is similar.
The higher values of chi-squared at higher latitudes is consistent with 
the greater deviation from a linear fit in the high-latitude 
panels of Figure~\ref{fig:fit_lin}. The systematic deviation of the near-surface
points contributes most to the chi-squared: these correspond to the 
high-degree modes and so motivates taking a closer look at those data.
(The scatter of the deepest points is large but less significant because of
the large error bars on those points.)

\begin{figure}[ht]
\resizebox{\hsize}{8cm}{\includegraphics[angle=90]
{{\fig}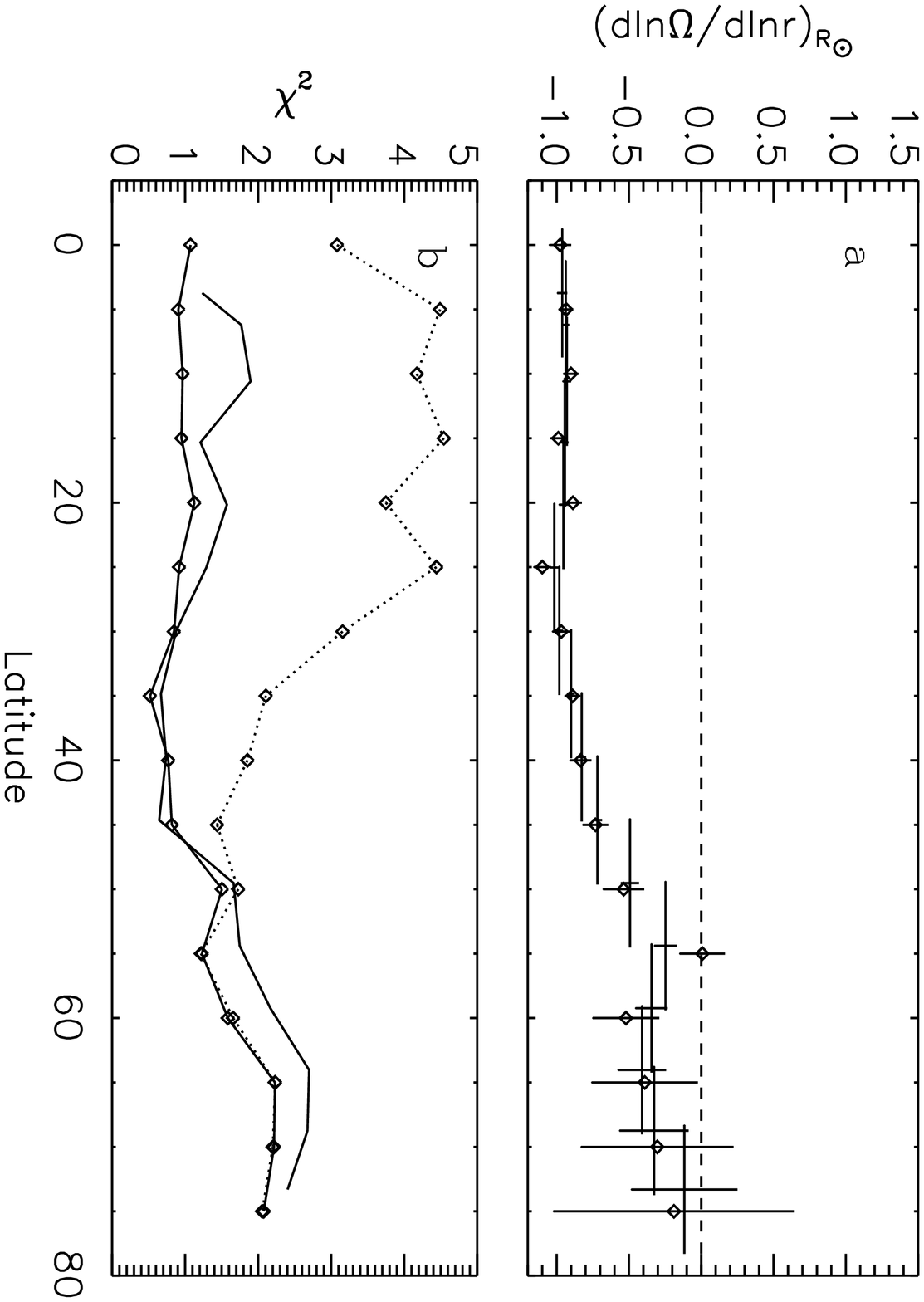}}
\caption{Similar to Figure~\ref{fig:ab-b-chi2} but using only modes $160\le l\le 250$. The radial gradient
of angular velocity remains negative even at latitudes above $55^\circ$. The dotted line on panel b 
shows, in the case of Equation~(\ref{eq:15D_bis}), the  $\chi^2$ values corresponding to a fit by a constant 
which is equivalent
of taking an average over $l$.} 
\label{fig:ab-chi2_160-250}
\end{figure}
We have therefore repeated our analysis but excluding those modes
of degree $l > 250$ and $l<160$. 
The resulting gradient and chi-squared are
shown in Figure~\ref{fig:ab-chi2_160-250}. Compared with the 
previous result (Figure~\ref{fig:ab-b-chi2}) the gradient is similar for
latitudes lower than 50 degrees. 
Now it is evident from Figure~\ref{fig:fit_lin} that, at high latitudes,
excluding the high-degree modes will tend to make the fitted gradient less
positive.
Indeed, we find that the gradient without the $l>250$ data remains slightly negative
up to about 75 degrees. Also, the values of 
chi-squared have been more than halved at high latitude, compared with our
previous linear fit to all the f-mode data (Figure~\ref{fig:ab-chi2_160-250}b).
The inferred low- and mid-latitude surface rate is barely affected
(compare the last two lines of Table~\ref{tab:table1}).

It is interesting also to compare the linear fit to the $l<250$ data with
a fit of a constant function to the same data: the constant fit is equivalent 
averaging the f-mode splittings over $l$ (see Table~\ref{tab:table1}). It is evident from 
Figure~\ref{fig:ab-chi2_160-250}b (dotted line) that this provides a very poor fit below 
about $55^\circ$: the data strongly favour the model with a linear 
depth-dependence 
there. At high latitudes, the linear fit selects only a very small gradient 
and so the two chi-squared functions are very similar: the data for
$l<250$ indicate that at high latitudes the gradient is small, in the 
range of depths spanned by their lower turning points.

\begin{figure}[ht]
\resizebox{\hsize}{!}{\includegraphics[angle=90]
{{\fig}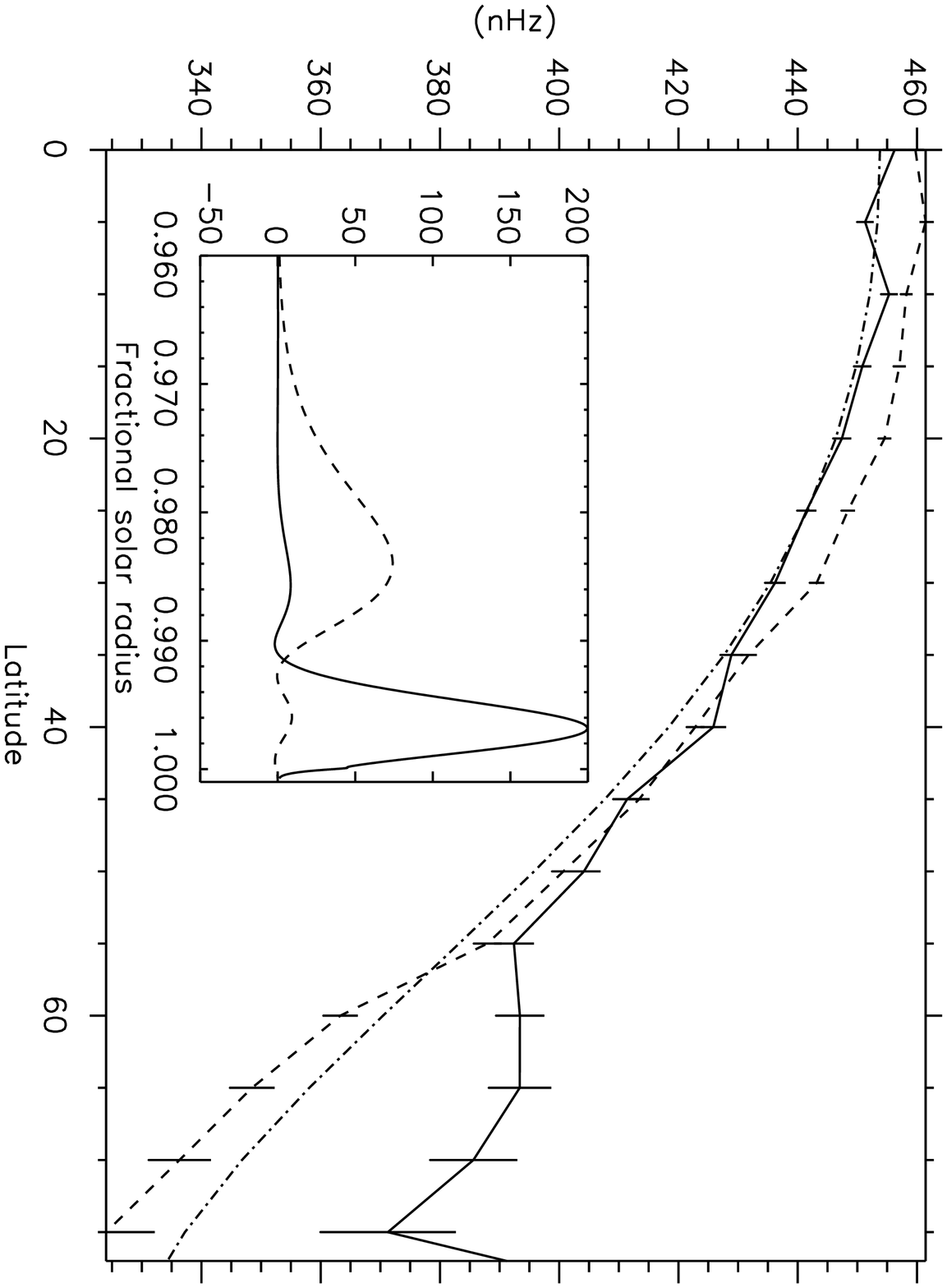}}
\caption{Rotation profiles as a function of latitude corresponding to depth
averaged shown in the sub panel. The dashed and full lines correspond respectively
to the shallower and deeper kernels  which have  respectively 
$0.986R_\odot$ and $0.997R_\odot$ as center of gravity. The dot-dashed line corresponds to the 
Snodgrass et al. (1984) plasma rotation rate. These results 
are obtained by  using all modes from $l=117$ to $l=300$.}
\label{fig:bg_r_kernels}
\end{figure}
If the data for $l>250$ are indeed reliable, then the discrepancy between
the results in Figures~\ref{fig:ab-b-chi2}a and \ref{fig:ab-chi2_160-250}a 
implies that
the model of rotation varying linearly with depth is not appropriate at
high latitudes and the extrapolation to the surface at those latitudes will
be unreliable. An alternative approach is to attempt to construct kernels
that are localized in depth using the Optimally Localized Averaging (OLA) 
kernel
in depth (cf. \opencite{CDST90})  in the manner of \cite{backus68}.
Such kernels at two selected
depths are shown in Figure~\ref{fig:bg_r_kernels}: they were constructed using
all the available f modes. It should be noted that the method succeeds in
producing kernels which are reasonably localized and which have their center of
gravity outside the range of abscissa values in Figure~\ref{fig:fit_lin},
that is, the method uses the mode sensitivities to extrapolate to greater
depths and closer to the surface. In particular, in the latter case one
expects that the increasing trend of values for the near-surface points in
Figure~\ref{fig:fit_lin}e,f  means that the near-surface
Backus-Gilbert inversion at those latitudes will have values higher than 
those seen in Figure~\ref{fig:fit_lin}. This is exactly what is found
(Figure~\ref{fig:bg_r_kernels}): the Backus-Gilbert inversion at high latitudes for
$r=0.986$ interestingly falls below the 3-term spectroscopic surface rate, but 
even more strikingly the corresponding near-surface result at $r=0.997$
lies above it by $2-4$ standard deviations. This is another way of 
demonstrating that the increasing values of the combined splittings for
$l>250$, if they are reliable, 
indicate a strongly positive gradient of rotation with radius in the 
rather superficial subsurface layers at high latitudes.

\begin{figure*}[ht]
\resizebox{\hsize}{8cm}{\includegraphics[angle=90]
{{\fig}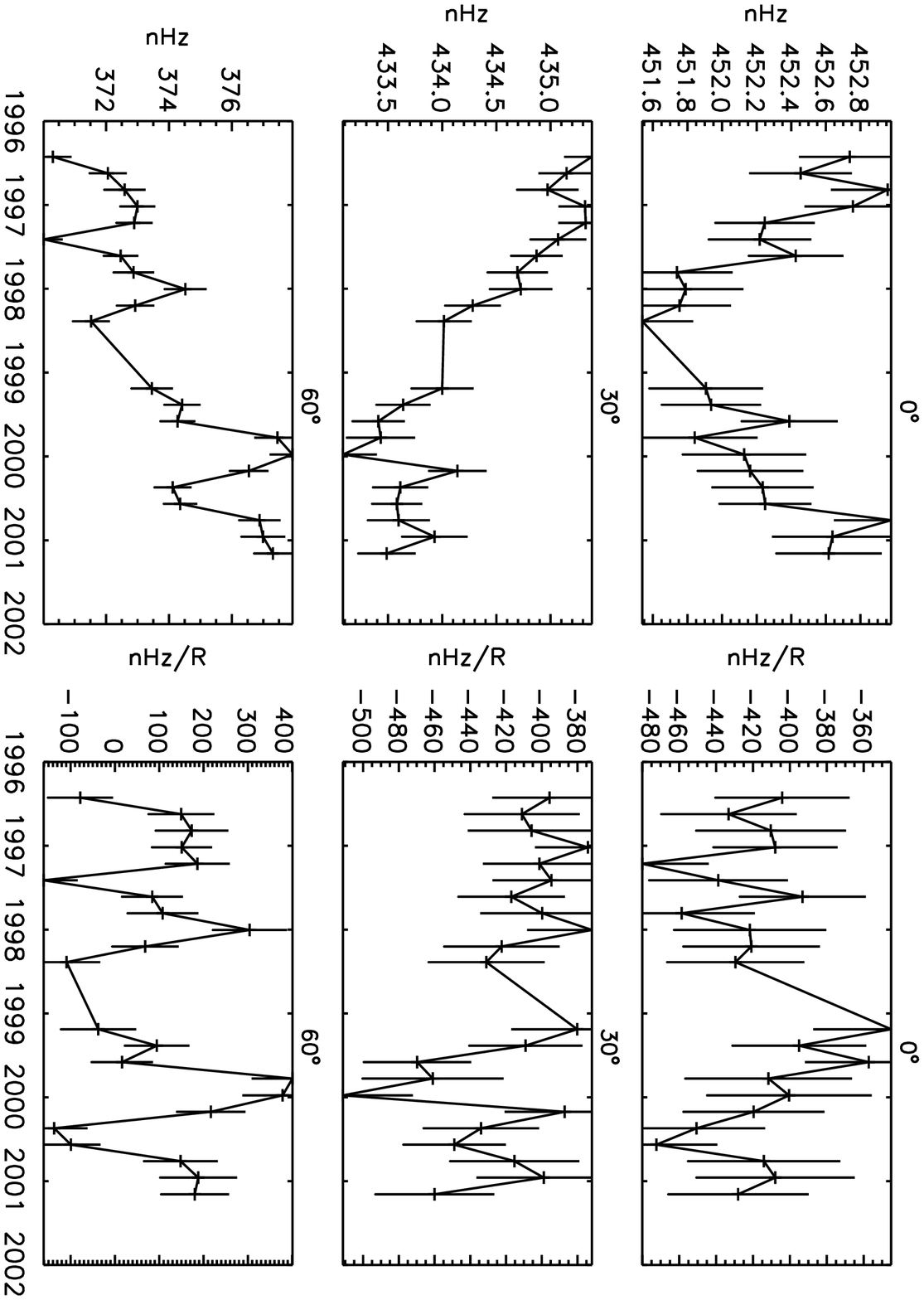}}
\caption{Intercept (left column) and slope (right column) 
of the linear fit Equation~(\ref{eq:lin_fit}) at
the equator, $30^\circ$ and $60^\circ$ of latitude (from top to bottom).}
\label{fig:time}
\end{figure*}
To look for possible temporal variations of the 
subsurface shear, we have analyzed each one of the 23 72-day datasets
individually in exactly the same manner as we analyzed the time-averaged
set (e.g. Figure~\ref{fig:fit_lin}), and 
derived an intercept value $\alpha(\mu_0; t)$
(corresponding to the surface rate at that location and epoch) and 
slope $\beta(\mu_0; t)$ from a linear fit to the combined splittings for
each latitudinal location $\mu_0$ and time $t$. The resulting estimated surface
rates and slopes at three latitudes (equator, $30^\circ$, $60^\circ$)
are shown in Figure~\ref{fig:time}. The large-scale variations in the surface
rate correspond very well to the migrating banded zonal flows (torsional
oscillations) measured by Schou (1999) and by Howe et al. (2000): the
equatorial surface rate starts high because of the tail-end of one 
migrating band of faster flow, then drops down and rises again towards
solar maximum as another band of faster flow reaches the equator: the latter
was at $30^\circ$ at the beginning of the cycle, hence the rate at that location
starts high and drops as the band migrates closer to the equator. The
$60^\circ$ rate rises as the high-latitude banded flow reported by
e.g. Schou (1999) strengthens towards solar maximum. 
The slope shows no significant corresponding variations, implying that
the torsional oscillations raise and lower the rotation rate across the
whole depth of the layer without changing the shear gradient.
There are indications of annual variations in the inferred values of the 
slope (most strikingly at $30^\circ$), which are almost certainly an
artifact: such artifacts can conceivably arise from annual variations in
SoHO's orbit. Other evidence for one-year artifacts in the f-mode data is 
presented by \inlinecite{antia_tenerife}. These should not affect the 
time-averaged values, however. There is no noticeable annual variation in
our inferred values of the surface rotation rate.

\section{Discussion}

We have used the depth and latitude variation in the sensitivities of the 
solar f modes to deduce the rotation profile in the subsurface shear
layer of the Sun in the outer 15 Mm of the solar interior. Our work 
differs from earlier seismic investigations. These were either based on 
the f modes but implicitly assumed a depth-independent model of the rotation
(e.g. \opencite{schou99}),
or used global inversions of p- and f-mode splittings and consequently may
suffer from any systematic difference between the p- and f-mode data
(e.g. \opencite{schou:mdi98}), or
used local helioseismic ring analysis
(e.g. \opencite{basu_ringdiag}; \opencite{haber_ringdiag}), which promises to be a powerful
diagnostic of near-surface flows and stratification but the sensitivity and
systematics of which are still under investigation (Hindman et al. 2001,
in preparation).  By using just the
splittings of the f modes, which are arguably the most straightforward 
helioseismic modes to interpret, we believe we are able to obtain
not only a simple but also a clean measure of the near-surface shear.
As with all inferences about rotation from global splittings, we 
note that only that component of rotation which is
symmetric about the equator is recovered. 

The most robust results concern the low-latitude shear. The average
gradient $\partial \ln\Omega / \partial \ln r$ (at constant latitude) in 
the outer 15 Mm is close to $-1$ and remarkably constant from the equator
to $30^\circ$ latitude. Between $30^\circ$ and $55^\circ$ latitude, the 
gradient is still negative but makes a steady transition to a small 
(absolute) value. All our analyses show this. The variation of rotation 
at these latitudes appears to be well described by a linear function of
depth, within the outer 15 Mm.

As discussed in the Introduction, if moving parcels of fluid were to 
conserve their specific angular momentum as they moved towards or away from 
the rotation axis, one
would find that the rotation rate varied as the inverse square of the 
distance from the axis of rotation, so at low latitudes one would have
that $\partial\ln \Omega / \partial\ln r \simeq -2$. In reality other 
effects such as diffusion will cause exchange of angular momentum between
parcels, so we may expect a logarithmic gradient somewhat smaller in 
magnitude than $-2$. A precise measurement of this value in the Sun provides
information about the relative effectiveness of competing mechanisms 
transporting angular momentum. Our finding is that at latitudes below $30^\circ$
the value of the logarithmic gradient is much closer to $-1$ than to $-2$.
In fact, this seems in reasonable agreement with the equatorial value 
found by DeRosa in numerical simulations of rotating compressible
convective fluid
in a thin shell representing the Sun between about $0.94R$ and $0.98R$
\cite{DeRosa_thesis,DeRosa01}. Also these simulations show a tendency for the gradient
to decrease in magnitude as one moves from equator to mid-latitudes, albeit
at lower latitudes than we find for the Sun. Although these simulations 
exclude for numerical reasons the near-surface layers that we are probing,
the qualitative agreement is nonetheless encouraging. 

At latitudes above $\sim 55^\circ$, the depth-averaged gradient over the 
layer appears to change sign with respect to the low-latitude shear, though
this is largely a consequence of the behaviour in the very near-surface 
layers (outer 5 Mm) which in turn is deduced from the splittings
of the highest-degree f modes. The gradient in
the range of depths $5-15$ Mm is small at these high latitudes; and such 
significant gradient $\partial\Omega/\partial r$
as does exist at high latitude (if any) is in the 
outer 5 Mm and predominantly positive. 
We note that, using a ring-analysis technique, \inlinecite{basu_ringdiag}
deduced a similar behaviour at high latitudes, finding 
a reversal of gradient in a zone above 0.994$R_\odot$.

Concerning the surface rotation rate itself, below $55^\circ$
our extrapolation of the rotation
rate to the surface is in satisfactory agreement with the directly 
measured spectroscopic surface rotation rate (cf. Table~\ref{tab:table1}).
Our inferred surface rate should be more accurate than one simply inferred
from the averaged f-mode splittings, because we take out the 
linear gradient with depth which undoubtedly exists at these latitudes:
this can make a difference of $\sim 5\,$nHz, even over the fairly small range 
of depths sampled by the observed f modes.

The seismically inferred surface rate at high latitudes is considerably
less secure. It has previously been reported from helioseismic 
investigations that the high-latitude surface rate is lower what one
would expect from a simple three-term extrapolation from lower latitudes
\cite{schou:mdi98,schou99}. 
Indeed it can be seen
from  panel f of Figure~\ref{fig:fit_lin} that many of the points 
fall below the extrapolated spectroscopic rate for that latitude, 
implying that the rotation rate
at {\sl some} depth is lower than the spectroscopic surface rate
one would infer from the values in Table~\ref{tab:table1}. The rather flat
plateau of values in those panels strongly suggests that the rotation 
rate at about 10-15 Mm depth is slower than the extrapolated spectroscopic
rate, which
is confirmed by our OLA inversion result at those depths. 
However,
the combined splittings at high degree are {\sl increasing} with $l$
and if taken at face value, as is done in our OLA inversion result for
$r=0.997R$, this behaviour implies that the very near-surface rotation rate 
is actually
higher than the spectroscopic rate. Thus the matter is still open.  Since
the quoted spectroscopic rate is principally an extrapolation of surface
observations at low- and mid-latitudes, the true rotation rate that 
would be determined by spectroscopy 
at high latitudes is
uncertain.  Direct spectroscopic determinations at high 
latitude would resolve the question.
The very high-degree splittings could contain some systematic errors, and 
if these affect the low-$m$ data the most (some evidence for such an 
effect for p modes at lower degrees is offered by the comparison of
GONG and MDI splittings by 
\inlinecite{Schou:comp01}), then the near-surface, 
high-latitude rotation rates inferred here could be erroneously high.
We hope that 
this possibility will shortly be addressed by independent determinations of
these splittings by the GONG experiment using the 
new higher-resolution GONG+ observations. 

\section{Conclusion}
Finally, to return again to our principal focus which is the shear gradient of
the near-surface rotation, we find that at low and mid-latitudes the 
gradient $\partial\Omega/\partial r$ in the
outer 15 Mm or so is close to $-1$ and is quite independent of latitude
below $30^\circ$; between $30^\circ$ and $\sim 50^\circ$ latitude, it is
still negative but makes a transition to small absolute value.
At higher
latitudes, the gradient in the bulk of the outer 15 Mm is probably small,
but if the highest-degree ($l>250$) data are to be believed there is a 
region of positive gradient in the outer 5 Mm at high latitudes, 
similar to what \inlinecite{basu_ringdiag} found from ring analysis. We find no
evidence for the gradient to vary with time: the torsional oscillation seems
to pass through without changing the shear gradient in the outer
15 Mm.

Interestingly, the most recent circulation-dominated
dynamo models \cite{dikpati_char99,Kuker01} are able to reproduce to some 
extent the equatorward migration patterns without invoking
any radial gradient of angular velocity at the surface. Such negative gradient 
at low latitude should 
however probably be taken into account because if it is associated
with a positive surface $\alpha$-effect, it will   
compete against the surface poleward circulation and contribute to producing
the equatorward migration of
magnetic patterns observed at the surface of the Sun.

\begin{acknowledgements}
We thank Marc DeRosa, Maussumi Dikpati, Fran\c cois Ligni\`eres and Peter Gilman
 for useful discussions. 
Prof. Juri Toomre and Dr. Steve Tomczyk are thanked
for hospitality at JILA and HAO respectively where part of this work was
carried out.
The work was supported by the UK Particle Physics \& Astronomy Research
Council through the award of grant PPA/A/S/2000/00171.
\end{acknowledgements}

\appendix
\section*{Derivation of f-mode 1.5D kernels}

The polynomial $\cal P$ used to describe the frequency splittings
 can be expressed in terms of the 
 Clebsh-Gordan coefficients $C_{j_1m_1j_2m_2}^{jm}$ (e.g. \opencite{Edmonds60})
 by:
\begin{equation}
{\cal P}_j^{(l)}(m)=\beta_j^l \ C_{l mj0}^{l m}; \ \ 
 \beta_j^l\equiv\frac{l\sqrt{(2l-j)!(2l+j+1)!}}{(2l)!\sqrt{2l+1}}. 
\end{equation}
The Gegenbauer polynomials used in Equation~(\ref{eq:omega}) are defined by  (e.g. \opencite{morse_feshbach53}):
\begin{equation}
T^1_{2j}(\mu)=\sqrt{\frac{4\pi}{4j+3}}\ \ \frac{\partial Y_{2j+1}^0(\theta,\phi)}{\partial \mu}\;.
\end{equation}
From \inlinecite{ritzwoller_lavely} we can deduce that
\begin{equation}\label{eq:15D_init}
2\pi a^{l}_{2j+1}= \frac{v_{2j+1}^l}{T_{2j}^1(0)} \int_0^1 \!\!K_j^l(r)
\Omega_{2j+1}(r)\dd r \;,
\end{equation}
where
 $v_{2j+1}^l\equiv L^2 C_{l 1 (2j+1) 0}^{l 1}/\beta_{2j+1}^l$, 
$L^2\equiv l(l+1)$ and
\begin{equation}\label{eq:kjl}
K_j^l(r)=\frac{\left(\xi_l^2+(L^2-1-j(2j+3))\eta_l^2-2\xi_l\eta_l\right)\rho r^2}{\int_0^1 \left(\xi_l^2+L^2\eta_l^2\right)\rho r^2 \dd r},
\end{equation}
$\xi_l$ and $\eta_l$ being respectively the radial and horizontal displacement 
eigenfunctions  which
are determined by solving the differential equations describing the motion of 
a self-graviting fluid body in a standard solar model (e.g. \opencite{unno89}) 
and $\rho$ is the density profile given by the model, all these being 
functions of the fractional solar radius $r$. 

Other expressions of practical interest can be found for $v_{2j+1}^l$ that are recalled
here for completeness. 
\inlinecite{pijpers97} established the recurrence relation
\begin{equation}
v_{2j+1}^l=\frac{(j-l)(2j+1)}{j(2l+2j+1)}v_{2j-1}^l\;;
\end{equation}
and \inlinecite{schou99} noticed that to a very good approximation
\begin{equation}\label{eq:v_alt}
v_{2j+1}^l/T_{2j}^1(0)\approx e^{-j(j+3/2)/l}.
\end{equation}

The f modes are horizontally propagating surface gravity waves for which
  the displacement 
eigenfunctions satisfy the following surface boundary condition under 
the Cowling approximation 
(e.g. \opencite{bertho_jcd91}): 
\begin{equation}\label{eq:boundary}
\eta_l(r)\approx\frac{g_s}{R_\odot w_l^2}\xi_l(r),
\end{equation} 
where $g_s=GM_\odot/R_\odot^2$ is the surface gravitational 
acceleration. Moreover, the 
angular frequencies $w_l=2\pi\nu_{l 0}$ of the f modes
follow asymptotically (for $l\rightarrow \infty$) the dispersion relation
$w_l^2\approx g_s L/R_\odot$. 
Therefore we have $\xi_l\approx L \eta_l$ and, from Equation~(\ref{eq:kjl}),
the rotational kernels associated with the f modes can 
be written as a function of the horizontal displacement only:
\begin{equation}\label{eq:rotkern}
K_j^l(r)\approx k_j^l K_h^l(r) \ \ \left\{\begin{array}{l}
 K_h^l(r)\equiv\frac{{\eta_l}(r)^2\rho(r)\ r^2}{\int_0^1{{\eta_l}(r)^2\rho(r)\ r^2\dd r}} \\
k_j^l\equiv1-\frac{1}{L}-\frac{1}{2L^2}(1+j(2j+3))\\
\end{array}\right.\;.
\end{equation}
Finally, Equation~(\ref{eq:15D}) is obtained by taking:
\begin{equation}
 u_{2j+1}^l \approx k_j^l\ e^{-j(j+3/2)/l} .
\end{equation}
We note that Equation~(\ref{eq:fundamental}) is obtained by using the fact that, 
in the approximation Equation~(\ref{eq:rotkern}) valid for f modes,
the rotational kernels depend on $j$ only by a multiplicative factor.
Taking instead $K_j^l\approx K_0^l $ for all $j$ as usually done for high degree modes
would also allow us to
write Equation~(\ref{eq:fundamental}) but the integrated difference 
$\int (K_j^l-K_0^l)\dd r$ would reach $2.2\%$ for $l=117$, $j=17$ whereas 
it remains negligible for all $l$ and $j$ in the case of the 
approximation used here.


\begin{thebibliography}{}

\bibitem[\protect\citeauthoryear{Antia et~al.}{2001}]{antia_tenerife}
Antia, H.~M., Basu, S., Pintar, J. and  Schou, J.: 2001,
\newblock in: A. Wilson (ed), {\em Helio- and Asteroseismology at the Dawn of the
Millennium}, ESA Publications Division, 
Noordwijk, The Netherlands, SP-464, p.~27.

\bibitem[\protect\citeauthoryear{Backus and Gilbert}{1968}]{backus68}
Backus, G.~E. and  Gilbert, J.~F.: 1968,
\newblock {\em Geophys. J. Roy. Astron. Soc.} {\bf 16}, 169.

\bibitem[\protect\citeauthoryear{{Basu} et~al.}{1999}]{basu_ringdiag}
{Basu}, S., {Antia}, H.~M. and {Tripathy}, S.~C.: 1999,
\newblock {\em Astrophys. J.} {\bf 512}, 458.

\bibitem[\protect\citeauthoryear{{Beck}}{2000}]{beck_review00}
{Beck}, J.~G.: 2000,
\newblock {\em Solar Phys.} {\bf 191}, 47.

\bibitem[\protect\citeauthoryear{{Beck} and {Schou}}{2000}]{beck00}
{Beck}, J.~G. and  {Schou}, J.: 2000,
\newblock {\em Solar Phys.} {\bf 193}, 333.

\bibitem[\protect\citeauthoryear{Berthomieu and
  Christensen-Dalsgaard}{1991}]{bertho_jcd91}
Berthomieu, G. and Christensen-Dalsgaard, J.: 1991,
\newblock in: A.~N. Cox, W.~C. Livingston and M. Matthews (eds.): {\em Solar
  Interior and Atmosphere}, The University of Arizona Press, Tucson, USA,  p.~412.

\bibitem[\protect\citeauthoryear{{Birch} and {Kosovichev}}{1998}]{birch98}
{Birch}, A.~C. and {Kosovichev}, A.~G.: 1998,
\newblock {\em Astrophys. J.} {\bf 503}, L187.

\bibitem[\protect\citeauthoryear{Brown et~al.}{1989}]{brown89}
{Brown}, T.~M., {Christensen-Dalsgaard}, J., {Dziembowski}, W.~A., 
        {Goode}, P., {Gough}, D.~O. and {Morrow}, C.~A.: 1989,
\newblock {\em Astrophys. J.} {\bf 343}, 526.

\bibitem[\protect\citeauthoryear{{Christensen-Dalsgaard} et~al.}{1990}]{CDST90}
{Christensen-Dalsgaard}, J., {Schou}, J. and {Thompson}, M.~J.: 1990,
\newblock {\em MNRAS} {\bf 242}, 353.

\bibitem[\protect\citeauthoryear{Collin et~al.}{1995}]{collin95}
Collin, B.,  Nesme-Ribes, E.,  Leroy, B.,   Meunier, N. and Sokoloff, D.: 1995,
\newblock {\em C.R. Acad. Sci. Paris} {\bf 321}, 111.

\bibitem[\protect\citeauthoryear{Corbard et~al.}{1997}]{corbard97}
Corbard, T., Berthomieu, G., Morel, P., Provost, J., Schou, J.  and  Tomczyk, S.:
  1997,
\newblock {\em Astron. Astrophys.} {\bf 324}, 298.

\bibitem[\protect\citeauthoryear{Corbard et~al.}{2001}]{corbard_tacho_rev}
Corbard, T., Jim\'enez-Reyes, S.~J.,  Tomczyk, S., Dikpati, M. and  Gilman P.:
  2001,
\newblock in: A. Wilson (ed), {\em Helio- and Asteroseismology at the Dawn of the
Millennium}, ESA Publications Division, 
Noordwijk, The  Netherlands, SP-464, p.~265.

\bibitem[\protect\citeauthoryear{DeRosa}{2001}]{DeRosa_thesis}
DeRosa, M.~L.: 2001,
\newblock Ph.D. thesis, University of Colorado, Boulder, USA.

\bibitem[\protect\citeauthoryear{{DeRosa} et~al.}{2001}]{DeRosa01}
{DeRosa}, M.~L., {Gilman}, P.  and {Toomre}, J.: 2001,
\newblock {\em Astrophys. J.}  {submitted}.

\bibitem[\protect\citeauthoryear{{Deubner} et~al.}{1979}]{Deubner79}
{Deubner}, F.-L.,  {Ulrich}, R.~K. and {Rhodes}, E.~J.: 1979,
\newblock {\em Astron. Astrophys.} {\bf 72}, 177.

\bibitem[\protect\citeauthoryear{{Dikpati} and
  {Charbonneau}}{1999}]{dikpati_char99}
{Dikpati}, M. and {Charbonneau}, P.: 1999,
\newblock {\em Astrophys. J.} {\bf 518}, 508.

\bibitem[\protect\citeauthoryear{{Duvall}}{1980}]{duvall80}
{Duvall}, T.~L.: 1980,
\newblock {\em Solar Phys.} {\bf 66}, 213.

\bibitem[\protect\citeauthoryear{Edmonds}{1960}]{Edmonds60}
Edmonds: 1960, {\em Angular Momentum in Quantum Mechanics}.
\newblock Princeton University Press, Princeton, New Jersey, USA. 

\bibitem[\protect\citeauthoryear{{Foukal}}{1972}]{foukal72}
{Foukal}, P.: 1972,
\newblock {\em Astrophys. J.} {\bf 173}, 439.

\bibitem[\protect\citeauthoryear{{Foukal}}{1977}]{foukal77}
{Foukal}, P.: 1977,
\newblock {\em Astrophys. J.} {\bf 218}, 539.

\bibitem[\protect\citeauthoryear{{Foukal} and
  {Jokipii}}{1975}]{foukal_jokipii75}
{Foukal}, P. and {Jokipii}, J.~R.: 1975,
\newblock {\em Astrophys. J.} {\bf 199}, L71.

\bibitem[\protect\citeauthoryear{{Gilman} and {Foukal}}{1979}]{gilman_foukal79}
{Gilman}, P.~A. and  {Foukal}, P.~V.: 1979,
\newblock {\em Astrophys. J.} {\bf 229}, 1179.

\bibitem[\protect\citeauthoryear{{Gilman} et~al.}{1989}]{gilman89}
{Gilman}, P.~A., {Morrow}, C.~A. and {Deluca}, E.~E.: 1989,
\newblock {\em Astrophys. J.} {\bf 338}, 528.

\bibitem[\protect\citeauthoryear{{Haber} et~al.}{2000}]{haber_ringdiag}
{Haber}, D.~A., {Hindman}, B.~W., {Toomre}, J., {Bogart}, R.~S., {Thompson}, M.~J.
  and {Hill}, F.: 2000,
\newblock {\em Solar Phys.} {\bf 192}, 335.

\bibitem[\protect\citeauthoryear{Hansen et~al.}{1977}]{hansen77}
Hansen, C.~J., Cox, J.P.  and  Van-Horn, H.~M.: 1977,
\newblock {\em Astrophys. J.} {\bf 217}, 151.

\bibitem[\protect\citeauthoryear{{Hathaway} et~al.}{1996}]{hathaway96}
{Hathaway}, D.~H., et~al.: 1996,
\newblock {\em Science} {\bf 272}, 1306.

\bibitem[\protect\citeauthoryear{{Howard}}{1984}]{howard84}
{Howard}, R.: 1984,
\newblock {\em ARA\&A} {\bf 22}, 131.

\bibitem[\protect\citeauthoryear{{Howard} and {Labonte}}{1980}]{howard80}
{Howard}, R. and  {Labonte}, B.~J.: 1980,
\newblock {\em Astrophys. J.} {\bf 239}, L33.

\bibitem[\protect\citeauthoryear{{Howe} et~al.}{2000}]{howe2000_band}
{Howe}, R., {Christensen-Dalsgaard}, J., et~al.: 2000,
\newblock {\em Astrophys. J.} {\bf 533}, L163.

\bibitem[\protect\citeauthoryear{{K{\" u}ker} et~al.}{2001}]{Kuker01}
{K{\" u}ker}, M.,  {R{\" u}diger}, G. and {Schultz}, M.: 2001,
\newblock {\em Astron. Astrophys.} {\bf 374}, 301.

\bibitem[\protect\citeauthoryear{{Kosovichev} and {Schou}}{1997}]{koso_schou97}
{Kosovichev}, A.~G. and {Schou}, J.: 1997,
\newblock {\em Astrophys. J.} {\bf 482}, L207.

\bibitem[\protect\citeauthoryear{{Leighton}}{1969}]{leighton69}
{Leighton}, R.~B.: 1969,
\newblock {\em Astrophys. J.} {\bf 156}, 1.

\bibitem[\protect\citeauthoryear{Morse and Feshbach}{1953}]{morse_feshbach53}
Morse, P. and  Feshbach, H.: 1953, {\em Methods of Theoretical Physics, Vol.~1},
\newblock McGraw-Hill, New York, USA.

\bibitem[\protect\citeauthoryear{{Parker}}{1993}]{parker93}
{Parker}, E.~N.: 1993,
\newblock {\em Astrophys. J.} {\bf 408}, 707.

\bibitem[\protect\citeauthoryear{{Pijpers}}{1997}]{pijpers97}
{Pijpers}, F.~P.: 1997,
\newblock {\em Astron. Astrophys.} {\bf 326}, 1235.

\bibitem[\protect\citeauthoryear{{Ritzwoller} and
  {Lavely}}{1991}]{ritzwoller_lavely}
{Ritzwoller}, M.~H. and  {Lavely}, E.~M.: 1991,
\newblock {\em Astrophys. J.} {\bf 369}, 557.

\bibitem[\protect\citeauthoryear{{Roberts} and {Stix}}{1972}]{roberts_stix72}
{Roberts}, P.~H. and {Stix}, M.: 1972,
\newblock {\em Astron. Astrophys.} {\bf 18}, 453.

\bibitem[\protect\citeauthoryear{Schou}{1992}]{schou_thesis}
Schou, J.: 1992,
\newblock Ph.D. thesis, Aarhus University, Aarhus, Denmark.

\bibitem[\protect\citeauthoryear{{Schou}}{1999}]{schou99}
{Schou}, J.: 1999,
\newblock {\em Astrophys. J.} {\bf 523}, L181.

\bibitem[\protect\citeauthoryear{{Schou} et~al.}{1994}]{schou94}
{Schou}, J., {Christensen-Dalsgaard}, J. and {Thompson}, M.~J.: 1994,
\newblock {\em Astrophys. J.} {\bf 433}, 389.

\bibitem[\protect\citeauthoryear{Schou et~al.}{1998}]{schou:mdi98}
Schou, J. et~al.: 1998,
\newblock {\em Astrophys. J.} {\bf 505}.

\bibitem[\protect\citeauthoryear{Schou et~al.}{2001}]{Schou:comp01}
Schou, J., Howe, R.,  et~al.: 2001,
\newblock {\em Astrophys. J.}
\newblock submitted. 

\bibitem[\protect\citeauthoryear{{Schroeter}}{1985}]{schroeter85}
{Schroeter}, E.~H.: 1985,
\newblock {\em Solar Phys.} {\bf 100}, 141.

\bibitem[\protect\citeauthoryear{{Snodgrass}}{1992}]{snodgrass92}
{Snodgrass}, H.~B.: 1992,
\newblock in: K.~L. Harvey (ed.), {\em ASP Conf. Ser. 27: The Solar Cycle}, 
ASP, San Francisco, USA, p.~205.

\bibitem[\protect\citeauthoryear{{Snodgrass} et~al.}{1984}]{snodgrass_HW84}
{Snodgrass}, H.~B., {Howard}, R. and  {Webster}, L.: 1984,
\newblock {\em Solar Phys.} {\bf 90}, 199.

\bibitem[\protect\citeauthoryear{{Snodgrass} and
  {Ulrich}}{1990}]{snodgrass_ulrich90}
{Snodgrass}, H.~B. and {Ulrich}, R.~K.: 1990,
\newblock {\em Astrophys. J.} {\bf 351}, 309.

\bibitem[\protect\citeauthoryear{Thompson et~al.}{1996}]{thompson:gong96}
Thompson, M.~J., Toomre, J., et~al.: 1996,
\newblock {\em Science} {\bf 272}, 1300.

\bibitem[\protect\citeauthoryear{{Unno} et~al.}{1989}]{unno89}
{Unno}, W., {Osaki}, Y., {Ando}, H., {Saio}, H.  and  {Shibahashi}, H.: 1989, {\em
  Nonradial oscillations of stars, 2nd ed.,}
\newblock University of Tokyo Press, Tokyo, Japan.

\end{thebibliography}

\end{article}
\end{document}